\documentclass[aps,pre,twocolumn,superscriptaddress]{revtex4}

\usepackage[utf8x]{inputenc}
\usepackage{ae}
\usepackage[T1]{fontenc}
\usepackage[english]{babel}
\usepackage{amsmath}
\usepackage{amsfonts}
\usepackage{amssymb}
\usepackage{array}
\usepackage{graphics}
\usepackage{graphicx}
\usepackage{epsfig}
\usepackage{latexsym}
\usepackage{textcomp}
\usepackage{gensymb}
\usepackage{verbatim}
\usepackage{subfigure}
\usepackage[retainorgcmds]{IEEEtrantools}
\newcommand{\rd}{\mathrm{d}}

\newcommand{\la}{\langle}
\newcommand{\ra}{\rangle}

\DeclareMathOperator{\erfc}{erfc}

\begin{document}

\title{Stochastic geometry and topology of non-Gaussian fields}

\author{T.H. Beuman}
 \affiliation{Instituut-Lorentz for Theoretical Physics, Leiden University, NL 2333 CA Leiden, The Netherlands}
\author{A.M. Turner}
 \affiliation{Institute for Theoretical Physics, Universiteit van Amsterdam, 1090 GL Amsterdam, The Netherlands}
\author{V. Vitelli}
 \email{vitelli@lorentz.leidenuniv.nl}
 \affiliation{Instituut-Lorentz for Theoretical Physics, Leiden University, NL 2333 CA Leiden, The Netherlands}

\begin{abstract}

Gaussian random fields pervade all areas of science. However, it is often the departures from Gaussianity that carry the crucial signature of the nonlinear mechanisms at the heart of diverse phenomena, ranging from structure formation in condensed matter and cosmology to biomedical imaging. The standard test of non-Gaussianity is to measure higher order correlation functions. In the present work, we take a different route. We show how geometric and topological properties of Gaussian fields, such as the statistics of extrema, are modified by the presence of a non-Gaussian perturbation. The resulting discrepancies give an independent way to detect and quantify non-Gaussianities. In our treatment, we consider both local and nonlocal mechanisms that generate non-Gaussian fields, both statically and dynamically through nonlinear diffusion.

\end{abstract}

\maketitle

Random fields pervade all areas of science. A disparate class of phenomena, ranging from the cosmic background radiation \cite{cite_Dodelson} and surface roughness \cite{cite_KPZ} to medical images of brain activity \cite{cite_Worsley} and optical speckle patterns \cite{cite_Flossmann}, produce data that can be regarded as random fields. The statistics of geometrical features of these fields, such as the density of extrema of various types, can be used to characterize them \cite{cite_Gruzberg, cite_Dennis1}. When the fields can be approximated as Gaussian fields, the physical meaning of these statistical properties is generally well understood \cite{cite_Longuet2,cite_Berry}: the statistics of extrema reflects the amount of field fluctuations at short distances.

Though analytical investigations are often restricted to Gaussian fields, phenomena described by nonlinear laws (such as the dynamics of inflation that produced the cosmic background radiation) produce non-Gaussian signals. Quite often, the observable signal is averaged over a large scale, producing approximately Gaussian statistics on account of the central limit theorem, and masking the nonlinearity.  Nevertheless, the surviving tiny departures from Gaussianity can carry the crucial signature of the nonlinear microscopic mechanisms at the heart of the phenomena.
As an illustration, consider a low resolution measurement of the spatial magnetization of a material well above the critical temperature. The magnetization fluctuates like a Gaussian random variable -- each region contains many domains oriented up or down in arbitrary proportion. However, a small non-Gaussian contribution remains, because there is a maximum possible magnetization per unit area which can be traced all the way down to the quantization of the spin of the electrons and hence the probability distribution cannot exhibit Gaussian tails.

In order to unveil such elusive effects, one needs an indicator that is sensitive to both \emph{short distances} and \emph{small signals}.

The most common tool used to probe the statistics of a random field is to measure its correlation functions. For example, the statistical properties of a random scalar field, $h(x,y)$, with Gaussian statistics, are entirely determined by its two-point correlation function $\langle h(x,y) h(x',y') \rangle$, and its higher-order correlation functions can be written simply as the sum of products of two-point correlation functions. The nonfactorizability of these higher-order correlation functions is one of the standard indicators of non-Gaussian statistics. 

\begin{figure}
\centering
\includegraphics{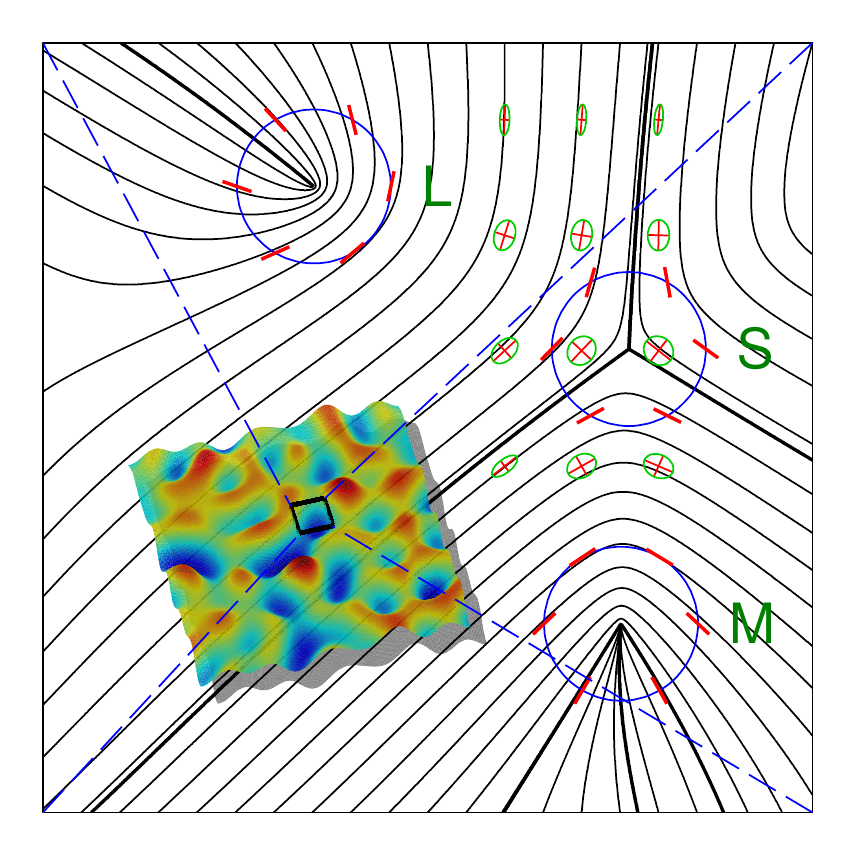}
\caption{The principal curvature of a surface. The major and minor axes of the small ellipses represent the direction and relative size of maximum and minimum curvature at the centre, equivalent to the direction of maximum and minimum polarization for an optical field. The curvature lines are always tangent to the direction of maximal curvature. At some points the curvature is the same along all directions (the equivalent polarization is circular); these are called umbilical points, of which there are three types, all shown in this image: on the top left is a lemon, on the top right a star and on the bottom right a monstar. The circles demonstrate their topological indices: $-1/2$ for a star, $+1/2$ for the other two. The lemon has one (locally) straight curvature line terminating at it (indicated with a thick line), the other two have three. The inset shows a computer-generated Gaussian surface with periodic boundary conditions, a small square of which served as the source of this picture.}
\label{fig_umbilics}
\end{figure}

Here we focus on a more geometric approach: view the scalar field as the height of a surface and study its random topography to infer the statistical properties of the signal (see inset of Fig.~\ref{fig_umbilics}). The densities of peaks and valleys, or of topological defects in the curvature lines known as umbilics (see Fig.~\ref{fig_umbilics}), are sensitive indicators of how jagged the height field is at \emph{short distances}. As we shall see, they provide an independent pipeline to detect non-Gaussianities, distinct from multiple-point correlation functions. This geometric approach has been applied successfully to track the power spectrum of a Gaussian field and it has been the subject of extensive theoretical and experimental studies \cite{cite_Dennis1, cite_Longuet2, cite_Berry, cite_Longuet1, cite_Dennis2}.

In this paper, we introduce the key physical concepts and mathematical techniques necessary to study the stochastic geometry of signals that can be described as a Gaussian random field plus a perturbation that we wish to track.
We first show how to treat non-Gaussianities within a \emph{local} approximation and calculate how the statistics of extrema change when a nonlinear transformation $F_{NL}(H_G)$ is applied locally to a Gaussian field $H_G$.
Then we consider the case of fields that cannot be probed directly, by calculating the statistics of \emph{umbilical points}, which are topological defects of the lines of principal curvature \cite{cite_Kamien}.
Finally, we turn to the class of nonlinear diffusion equations \cite{cite_KPZ} and go beyond the local approximation, by considering the effects of spatial gradients that couple values of the field at different locations. As an illustration, we solve explicitly for the nonlocal non-Gaussianities generated dynamically by the deterministic KPZ equation which models surface growth \cite{cite_KPZ}.

\section{Critical points} 
\label{sec_crit}

To gain some insights into the physical mechanisms that generate non-Gaussianities, consider first how an isotropic Gaussian field $H_G(\vec{r})$ arises from the random superposition of waves (or equivalently, Fourier modes) 
\begin{equation}
H_G(\vec{r}) = \sum_{\vec{k}} A(k) \cos( \vec{k} \cdot \vec{r} + \phi_{\vec{k}} ),
\label{eq1}
\end{equation}
\noindent with an amplitude spectrum $A(k)$\footnote{The power spectrum $A(k)^2$ is the Fourier transform of the two-point correlation function.} that depends only on the magnitude of the wave vectors, $k = |\vec{k}|$. The phases $\phi_{\vec{k}}$ are uncorrelated random variables uniformly distributed in the range $[0, 2\pi]$. The statistical properties of the Gaussian field $H_G(\vec{r})$ are entirely encoded by the function $A(k)$, or equivalently, its moments $K_{2n}=\sum_{\vec{k}} k^{2n} \frac12 A(k)^2$. 

The most basic difference between Gaussian and non-Gaussian variables is that Gaussian ones are always symmetric about their mean. As a consequence, irrespective of its power spectrum, a Gaussian field has equal densities of maxima and minima. Hence, a nonvanishing imbalance $\Delta n$ between these two types of extrema serves as a probe to detect and quantify the non-Gaussian component of a signal, provided that it can be measured directly.

For example, consider the primordial curvature perturbation field, $\Phi$, a nearly Gaussian field of central interest to modern cosmological studies \cite{cite_Dodelson}. Within a \emph{local} approximation, the primordial field is obtained from a Gaussian field $\Phi_G$ via a nonlinear relation $\Phi = \Phi_G + f_{nl} \Phi_G ^2 + g_{nl} \Phi_G ^3$. Determining the parameters $f_{nl}$ and $g_{nl}$ is one of the central tasks in the study of cosmological non-Gaussianities. As we shall see, the quadratic coefficient can be determined from the imbalance $\Delta n$ between maxima and minima of $\Phi$.

The imbalance can be derived in the more general context of a non-Gaussian field $h$ that is obtained from a Gaussian field $H_G$ via any nonlinear deformation $h=F_{NL}(H_G)$. If $F_{NL}$ is a monotonic function, the maxima and minima do not change -- only a nonmonotonic behavior of $F_{NL}$ can alter this balance. 

The critical points of $h$ are given by $ \vec{\nabla} h = F_{NL}'(H_G) \vec{\nabla} H_G = 0$, where the dash indicates the derivative of $F_{NL}(H_G)$ with respect to $H_G$. This condition shows that $h$ and $H_G$
have the same critical points. Note however that, if $F_{NL}'(H_G(\vec{r_0})) < 0$ at a critical point $\vec{r}_0$, then a maximum (minimum) at $H_G(\vec{r}_0)$ will be turned into a minimum (maximum) at $h(\vec{r}_0)$. The number of saddle points does not change because of topological constraints (it is equal to the invariant sum of maxima and minima).

If the transformation has a bias towards converting minima into maxima, then $h$ will have more maxima than minima; for example, $h=H_G+\varepsilon H_G^2$ reverses its slope at sufficiently negative values of $H_G$, which are most likely to be minima. Following this logic, the first step toward calculating the imbalance between maxima and minima is to determine the probability $g(z)$ that $H_G(\vec{r_0}) = z$ for a minimum $\vec{r_0}$ of $H_G$. The symmetry properties of $H_G$ imply that the analogous probability distribution for maxima is $g(-z)$. 

The fraction of minima of $H_G$ that become maxima of $h$ is obtained by integrating $g(z)$ over the range of $z$ for which $F_{NL}'(z) < 0$. Likewise, the fraction of maxima of $H_G$ that are turned into minima is given by the integral of $g(-z)$ over the same range. The overall imbalance in the densities of the maxima and minima of $h$ can be readily obtained by adding these opposite contributions. The result reads

\begin{equation}
\begin{split}
  \Delta n	& \equiv \frac{n_{\mathrm{max}} - n_{\mathrm{min}}}{n_{\mathrm{max}} + n_{\mathrm{min}}} \\
  		& = \int_{z:F_{NL}'(z)<0} \! \rd z \, \big( g(z) - g(-z) \big).
\end{split}
\label{eq_maxmin}
\end{equation}

\begin{figure}
\centering
\includegraphics{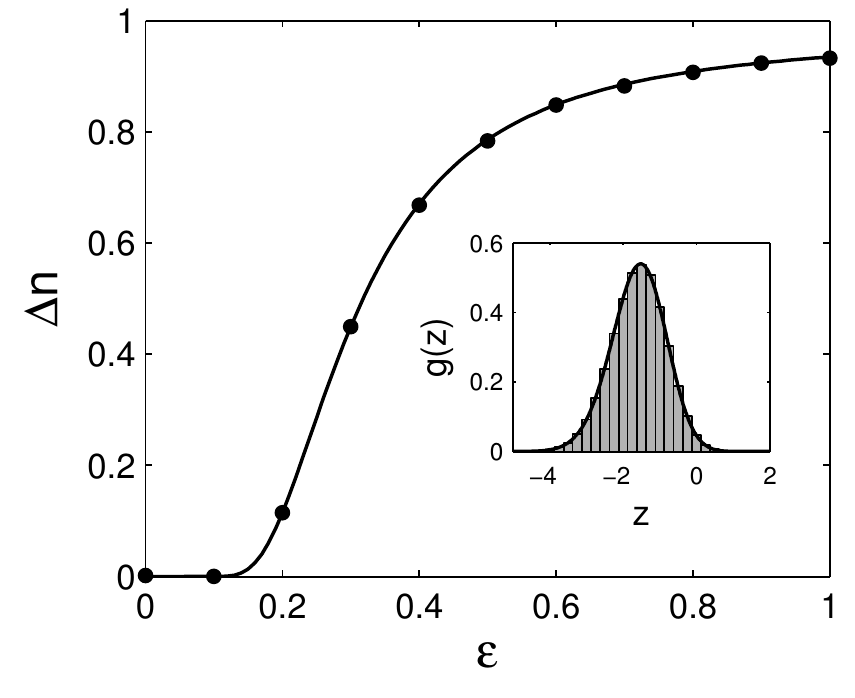}
\caption{The relative difference $\Delta n$ between the densities of maxima and minima of $h = H_G + \varepsilon H_G^2$, where $H_G$ is a Gaussian field with $\lambda = 3/4$, as a function of $\varepsilon$. The data points are results from computer-generated fields, the solid curve is the theoretical result (Eq.~\eqref{eq_maxmin}). The inset shows the corresponding distribution of minima $g(z)$ (which forms the basis of our theoretical result), both the theoretical curve and a histogram of data gathered from computer-generated fields.}
\label{fig_maxmin}
\end{figure}

For two dimensions, the exact analytical expression for $g(z)$ is explicitly derived in Appendix A -- it depends only on the moments $K_0$, $K_2$ and $K_4$. Rescaling $h(\vec{r}) \rightarrow h(a\vec{r})$ does not affect the function $g(z)$, since it increases the density of maxima with any value of $H_G$ by the same proportion. Hence, only $K_0=\langle H_G^2\rangle$ (which sets the scale of the distribution) and the dimensionless parameter $\lambda \equiv \frac{K_2^2}{K_0K_4}$ can enter in the expression for $g(z)$ (see plot in the inset of Fig.~\ref{fig_maxmin}).

As an illustration, apply Eq.~\eqref{eq_maxmin} to the perturbed Gaussian field $h = H_G + \varepsilon H_G^2$. Figure~\ref{fig_maxmin} shows our theoretical formula as a continuous line, validated by numerical data (dots) obtained from computer-generated random surfaces with amplitude spectrum $A(k) \sim \theta(k_D-k)$, having $\lambda = \tfrac34$.

The imbalance between maxima and minima is particularly useful to track large deviations from Gaussianity. This can be seen explicitly for the quadratic perturbation considered above. Note that, since $H_G$ itself has an equal number of maxima and minima, the imbalance $\Delta n$ in $h$ is only created when one of these critical points is inverted. Whether $H_G$ has a high likelihood of having a negative $F_{NL}'(H_G)=1+2\varepsilon H_G$ is controlled by $\varepsilon \sqrt{\langle H_G^2\rangle}$. The probability of $|2 \varepsilon H_G|$ exceeding $1$ is exponentially small. Thus, $\Delta n$ rises roughly as $e^{-c/({\varepsilon^2}\langle H_G^2\rangle)}$, where $c \approx \frac{1}{8}$. Note that our approach in deriving Eq.~\eqref{eq_maxmin} and $g(z)$ is nonperturbative and hence capable of describing large deviations from Gaussianity (for the specific type of deviation considered here), as demonstrated by the agreement between our formula and numerics over the entire $\varepsilon$-range probed in Fig.~\ref{fig_maxmin}.

\section{Umbilical points}

The near-Gaussian fields under investigation are not always directly accessible experimentally. For example, the mass distribution along the line of sight responsible for weak gravitational lensing is mostly composed of dark matter and hence it cannot be detected directly \cite{cite_Hoekstra}. If the projected gravitational potential over a flat patch of the sky is taken to be the height of a 2D surface \cite{cite_Vitelli}, the \emph{measurable} shear field, is given by the lines of principal curvature \cite{cite_Kamien}, as shown in Fig.~\ref{fig_umbilics}. At some special points called umbilics, the curvature is equal in all directions, so the shear field cannot be defined and it must vanish. More precisely, a point $\vec{r} = \{x, y\}$ on a surface with height function $h(\vec{r})$ is an umbilic if the second derivatives satisfy the two conditions $h_{xx}(\vec{r}) = h_{yy}(\vec{r})$ and $h_{xy}(\vec{r}) = 0$. The ratio between different types of umbilical points (which is a universal number for an isotropic Gaussian field) serves as an indicator of non-Gaussianities in lieu of the extrema which cannot be detected. A similar reasoning can be applied to study polarization singularities in the cosmic microwave background \cite{cite_Dennis3} and topological defects in a nematic or superfluid near criticality \cite{cite_Halperin,cite_Liu}.

Inspection of Fig.~\ref{fig_umbilics} reveals that there are three types of umbilics: lemons, monstars and stars. Note that these umbilics are topological defects in the curvature-line field. The topological index of any umbilic is equal to plus (minus) $1/2$, if the curvature-line field rotates clockwise (counter-clockwise) by an angle $\pi$, along any closed path encircling that umbilic only in the clockwise direction.

A star has three curvature lines terminating at it and a topological index of $-\tfrac12$. A lemon has only one line and index $+\tfrac12$. A monstar has index $+\tfrac12$, like a lemon, but three lines terminating at it, like a star.
A striking feature of isotropic Gaussian fields is that the monstar fraction, the relative density of monstars with respect to all umbilics, equals $\alpha_M = \frac12 - \frac1{\sqrt5} = 0.053$; this is a universal number independent of the power spectrum \cite{cite_Berry, cite_Dennis2}. Any deviation from this special value is therefore a sure sign of non-Gaussian effects.

We will again consider a height field that is a nonlinear function of a Gaussian, $h = F_{NL}(H_G)$. In this general case, we were not able to find an exact result as we were for the extrema. We shall therefore assume that the nonlinear contribution is small, such that $h=H_G+\varepsilon f(H_G)$, with $f$ a nonlinear perturbation and $\varepsilon \ll 1$ a small parameter controlling the size of the nonlinearity. We can express $\alpha_M$ in terms of the joint probability distribution $p$ of the second and third derivatives of $h$. To calculate $p$, we use the property that a probability distribution is determined by its moments of all orders (if the distribution is well-behaved). This is most easily accomplished by calculating the generating function of the distribution, $\chi$ -- its logarithm can be directly expressed in terms of the \emph{cumulants} $C_n$

\begin{equation}
 \log \chi(\lambda_1,\dots,\lambda_n)  =  \sum_{m=1}^\infty \frac{i^m}{m!} \!\! \sum_{j_1, \ldots, j_m} \!\!\!\! C_m(\xi_{j_1}, \ldots, \xi_{j_m}) \lambda_{j_1} \ldots \lambda_{j_m},
\label{cum} 
\end{equation}

\noindent where $\xi_1, \ldots, \xi_n$ are the stochastic variables given by the spatial derivatives of $h$. The probability distribution can then be obtained from the generating function by taking the inverse Fourier transform with respect to $\lambda_1, \ldots, \lambda_n$.
The cumulants can be written in terms of expectation values, e.g.\ $C_2(\xi_1, \xi_2) = \la \xi_1 \xi_2 \ra - \la \xi_1 \ra \la \xi_2 \ra$. In this context, these expectation values are called \emph{moments}, which are not to be confused with the moments $K_{2n}$ defined before. 

For Gaussian variables, only the second-order cumulants are nonzero. This gives rise to a generating function of the form
\begin{equation}
 \log \chi  =  -\frac{1}{2} \sum_{ij} C_2(\xi_i, \xi_j) \lambda_i \lambda_j.
\label{eq_chi}
\end{equation}
\noindent One can easily check that the inverse Fourier transform of $\chi$ in Eq.~\eqref{eq_chi} precisely yields the probability distribution for a set of correlated Gaussian variables (assuming $\la \xi_i \ra = 0$), see Eq.~\eqref{eq_gsn_jpd}. More generally, by determining all the cumulants, one can construct the generating function and from that the probability distribution. We shall derive the monstar fraction up to first order in the perturbation $\varepsilon f(H_G)$ only; consistently we need to determine all the cumulants up to first order only.

The monstar fraction (even at order $\varepsilon$) could in principle depend on $f$ in a complicated way, if $f$ is an arbitrary nonlinear function.
In fact, quadratic terms in the function $f$ produce degree $3$ cumulants in the distribution function of the field $h$ (i.e., skewness), cubic terms produce kurtosis (degree $4$ cumulants) and in general degree $n$ terms in $f$ produce degree $n+1$ cumulants. However, the monstar fraction can be determined from just the distribution of a few derivatives of $h$, whose cumulants vanish beyond the fourth order due to symmetry, as shown in the Supporting Information. Consequently, the final result for the monstar fraction depends only on a single parameter, $\la f'''(H_G)\ra=\int_{-\infty}^\infty f'''(u) e^{\frac{-u^2}{2K_0}}\frac{du}{\sqrt{2\pi K_0}}$, where the primes indicate derivatives with respect to $H_G$.

\begin{figure}
\centering
\includegraphics{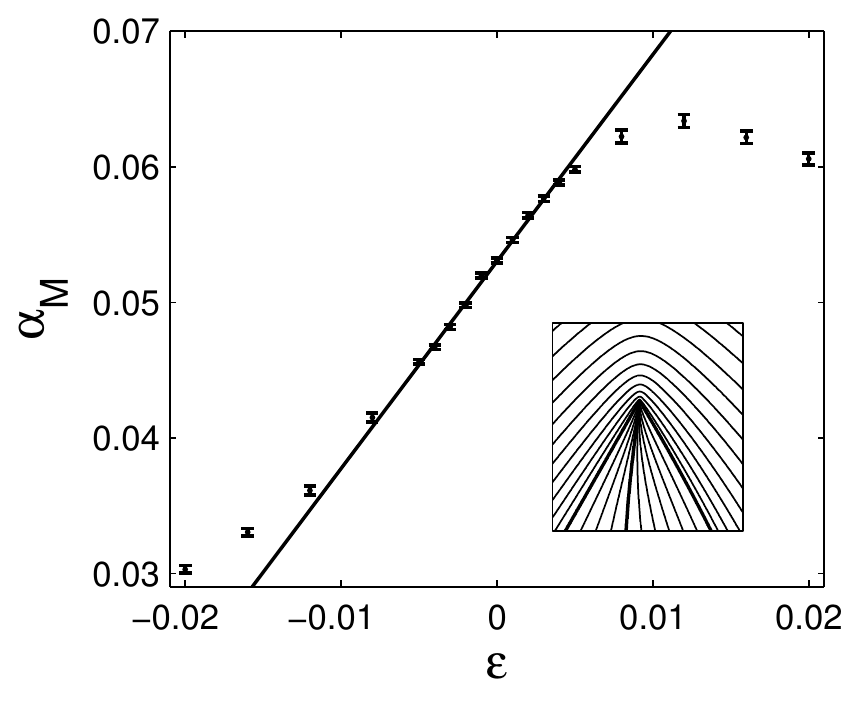}
\caption{The fraction of monstars (see inset) $\alpha_M$ of $h = H_G + \varepsilon H_G^3$, where $H_G$ is a Gaussian field with $\mu = 16/27$, as a function of $\varepsilon$. The data points are results from computer-generated fields, the solid line is the theoretical first-order result (Eq.~\eqref{eq_mfr}). At $\varepsilon = 0$ we retrieve the universal fraction $\alpha_M = 1/2 - 1/\sqrt5 = 0.053$, valid for any isotropic Gaussian field.}
\label{fig_mfr}
\end{figure}

\begin{table}
\centering
\caption{All nonzero cumulants}
\renewcommand{\arraystretch}{1.2}
\begin{tabular*}{\hsize}{@{\extracolsep{\fill}}@{\hspace{0.5cm}}ll@{\hspace{0.5cm}}}
  \hline
  $C_2(h_{zz}, h_{z^*z^*})$						&	$\sigma(1 + 2 \la f'(H) \ra)$	\cr
  $C_2(h_{zzz}, h_{z^*z^*z^*})$					&	$\tau(1 + 2 \la f'(H) \ra)$	\cr
  $C_2(h_{zzz^*}, h_{zz^*z^*})$					&	$\tau(1 + 2 \la f'(H) \ra)$	\cr
  $C_3(h_{zz}, h_{zzz^*}, h_{z^*z^*z^*})$ \& conj.			&	$-3 \sigma^2 \la f''(H) \ra$	\cr
  $C_4(h_{zzz^*}, h_{zzz^*}, h_{zz^*z^*}, h_{zz^*z^*})$		&	$-8 \sigma^3 \la f'''(H) \ra$	\cr
  $C_4(h_{zzz}, h_{zz^*z^*}, h_{zz^*z^*}, h_{zz^*z^*})$ \& conj.	&	$-6 \sigma^3 \la f'''(H) \ra$	\cr
  \hline
\end{tabular*}
\label{tbl_cumulants}
\end{table}

The calculation can be briefly summarized as follows. The monstar fraction is related to the distribution function of some of the second and third derivatives of $h$, which we write in terms of complex coordinates $z=x+iy$ and $z^*$: $h_{zz}$, $h_{zzz}$, and $h_{zzz^*}$. The definition of an umbilic point becomes $h_{zz} = 0$, where $h_{zz}$ is now complex. All the cumulants of these variables and their conjugates may now be calculated (up to order $\varepsilon$). The complex coordinates allow for optimal usage of rotational and translational symmetry. Only a few of the cumulants are nonzero and these are evaluated in Table~\ref{tbl_cumulants}.
With the aid of these cumulants, the generating function can be constructed to first order using Eq.~\eqref{cum}. Taking the Fourier transform leads to the probability distribution $p(h_{zz},h_{zzz},h_{zzz^*})$; the explicit form is rather long, but it is basically a Gaussian perturbed by cubic and quartic terms in $h$ and its derivatives \cite{cite_longpaper2}. To obtain $\alpha_M$, we then have to set $h_{zz} = 0$ and integrate over $h_{zzz}$ and $h_{zzz^*}$ (taking care to include the appropriate Jacobian factor). Integrating over \emph{all} of $\mathbb{C}^2$ gives the total density of umbilical points, while the density of monstars is obtained by integrating over a specific range, which can be found in Appendix B. The monstar fraction is then the ratio of these two densities. The resulting deviation from $\alpha_M = 0.053$ is

\begin{equation}
\Delta \alpha_M = 0.429 \mu \la f'''(H_G) \ra \varepsilon,
\label{eq_mfr}
\end{equation}

\noindent where $\mu \equiv K_4^3/K_6^2$. When applied to the local modal of the primordial field $\Phi$ described before, $\Delta \alpha_M$ in Eq.~\eqref{eq_mfr} depends only on the cubic coefficient $g_{nl}$ and not on $f_{nl}$.

Hence, the leading order perturbation that alters the monstar fraction is $f(H_G) = H_G^3$, for which $\la f'''(H_G) \ra = 6$.
Note that this perturbation, like any odd and/or monotonic function of $H_G$, does not have an effect on the density of maxima and minima. 
Figure~\ref{fig_mfr} shows $\alpha_M$, as determined by Eq.~\eqref{eq_mfr} (continuous line), together with data from computer simulations (symbols): the agreement between theory and numerics is very good in the linear regime. The spectrum used was again $A(k) \sim \theta(k_D-k)$, for which $\mu = \tfrac{16}{27}$. Note that the monstar fraction is very sensitive to a small non-Gaussianity, with a $20\%$ change when $\varepsilon$ is just $0.01$. For larger values of $\varepsilon$, nonlinear effects become important and prevent $\alpha_M$ from becoming negative. In this regime, our approximate result must break down.

\begin{figure}
\centering
\includegraphics{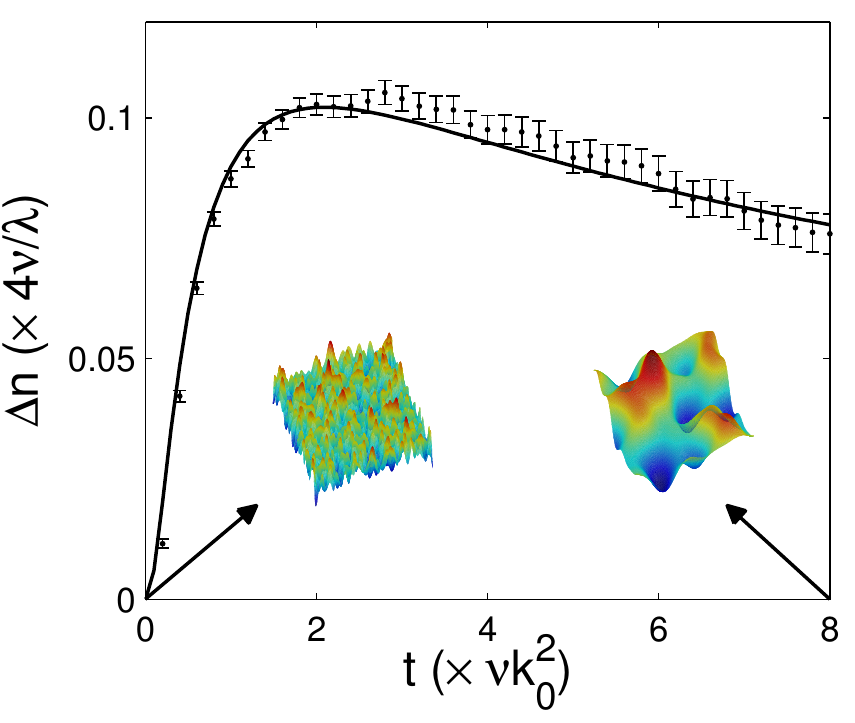}
\caption{The imbalance between maxima and minima $\Delta n$, as a function of time, for an initially Gaussian field evolving according to the deterministic KPZ equation (Eq.~\eqref{eq_kpz}). At $t=0$, the Gaussian field was taken to have a Gaussian power spectrum ($A(k)^2 \sim \exp(-k^2/2k_0^2)$); see inset on the left. As time evolves, the surface becomes smoother (see inset on the right), decreasing the densities of maxima and minima, but also creating an imbalance between the two. The data points stem from simulations, for which $\lambda/4\nu = 0.1$ was used. The solid curve is the theoretical result. }
\label{fig_kpz}
\end{figure}

\section{Nonlocal model and evolution equations}

In section \ref{sec_crit}, we presented an exact expression for the imbalance between maxima and minima for a local perturbation of a Gaussian random field. This simple class of models may describe the local evolution of a system that starts out with a Gaussian distribution, as for instance the growth of a population of cells that are initially distributed on a dish and then divide without any significant migration from one region to another.

However, many dynamical systems evolve in a nonlocal, nonlinear way. Non-Gaussianities are generated dynamically from the nonlinear equations of motion that the field $h(\vec{r},t)$ obeys, even if the initial condition $h(\vec{r},0)=H_G(\vec{r})$ is Gaussian. Unlike the case of local evolution, the imbalance between maxima and minima will now exhibit a power law increase, as the nonlinear perturbation grows.

A broad class of nonlinear diffusion equations describes the necessary mixing between regions. Examples include several models of structure formation in both condensed matter \cite{cite_Chaikin} and cosmology \cite{cite_Dodelson}, the Cahn-Hilliard equation for the development of order after a phase transition \cite{cite_Bray} and simplified models of surface growth \cite{cite_KPZ}. We will focus on the last of
these, the deterministic KPZ equation \cite{cite_KPZ} which models the height evolution of a substrate as atoms accumulate on it:

\begin{equation}
\frac{\partial h(\vec{r}, t)}{\partial t}  =  \nu \nabla^2 h(\vec{r}, t) + \frac{\lambda}{2} (\nabla h(\vec{r}, t))^2.
\label{eq_kpz}
\end{equation}
This equation is arrived at in the following way: to first order, the surface simply grows at a constant rate. This constant rate does not appear in the equation, however, since it is simply subtracted out; the two terms on the right-hand side are the additional effects. The first term describes the diffusion of particles along the surface, and the second nonlinear term describes approximately how the growth-rate varies with the local slope. The surface is assumed to grow at a constant rate perpendicular to itself, but since the height is measured vertically, $\dot{h}$ depends on the slope: this gives rise to the term quadratic in $\nabla h$.

Our approach can be applied to other nonlinear diffusion problems well beyond surface-growth dynamics, when a different choice for the quadratic term is made. For example, if the term quadratic in the gradient in Eq.~\eqref{eq_kpz} is substituted by a term quadratic in the field, $-h^2$, one obtains the Fisher equation which describes the growth and saturation of a population. A third possibility is to consider a mixed term $h \nabla h$ which gives the Burgers' equation governing shock dynamics and traffic flow. In all of these cases, we can study the time evolution of an initially Gaussian $h(\vec{r})$ profile. Upon setting the coefficient of the nonlinear term $\lambda$ equal to zero, we always retrieve the heat equation, which preserves the Gaussianity of $h$ for all later times. On the other hand, for $\lambda \neq 0$, $h$ becomes non-Gaussian. 

For concreteness, we discuss how an imbalance between maxima and minima is generated by nonlocal non-Gaussianities in the context of the KPZ equation (Eq.~\eqref{eq_kpz}). The nonlinear term breaks the symmetry between positive and negative values of $h$ which is a necessary condition to generate an imbalance. Note however that, in the case of a local evolution, this imbalance was exponentially small because the local evolution cannot create new extrema -- it can only convert a maximum into a minimum whenever $h$ happens to have a sufficiently large fluctuation. It is the presence of the diffusion term that is able to create new maxima and minima, even though, on its own, it would not be able to generate any \emph{imbalance}, because of the symmetry $h\rightarrow -h$. The two terms on the right-hand side of Eq.~\eqref{eq_kpz} conspire together to change the number of maxima and minima asymmetrically.

Since the imbalance between maxima and minima will now have a contribution perturbative in $\lambda$, we will determine a general expression for a field that is close to being Gaussian. The distribution of maxima and minima can be calculated from the joint distribution of $h_{z}$, $h_{zz}$ and $h_{zz^*}$, which can again be determined given the cumulants of these variables. We will assume that the third order cumulants are of order $\lambda$, while higher order cumulants are of higher order. (The formula is valid for the KPZ problem, since the assumption on the cumulants is satisfied for any function that evolves nonlinearly from an initially Gaussian signal, as long as the quadratic terms have coefficients of order $\lambda$, and for times that are not too large.) The relevant cumulants up to third order are named in Table~\ref{tbl_correlations}. 

\begin{table}
\centering
\caption{All second and third order cumulants}
\renewcommand{\arraystretch}{1.2}
\begin{tabular*}{\hsize}{@{\extracolsep{\fill}}@{\hspace{1.5cm}}ll@{\hspace{1.5cm}}}
  \hline
  $\sigma = \langle |h_z|^2 \rangle\ $		&	$\ \beta = \langle |h_z^2|h_{zz^*} \rangle$	\cr
  $\alpha = \langle |h_{zz}|^2 \rangle\ $		&	$\ \gamma = \langle h_{zz^*}^3 \rangle$		\cr
  							&	$\ \delta = \langle |h_{zz}|^2h_{zz^*} \rangle$	\cr
  \hline
\end{tabular*}
\label{tbl_correlations}
\end{table}

The expression for the imbalance $\Delta n$ is derived by writing the joint probability distribution in terms of the cumulants, as before. We then set $h_z=0$ and integrate over the range of $h_{zz}$ and $h_{zz^*}$ that defines the maxima and minima, to find
\begin{equation}
\Delta n  =  \sqrt{ \frac{6}{\pi \alpha} } \bigg( \frac43 \frac{\beta}{\sigma} + \frac49 \frac{\delta}{\alpha} - \frac{10}{27} \frac{\gamma}{\alpha} \bigg).
\label{eq_general}
\end{equation}

Equation~\eqref{eq_general} is a general result. In order to apply it to the KPZ equation, we first change variables to
\begin{eqnarray}
u(\vec{r}, t)  &=&  \frac{2\nu}{\lambda} \left[ \exp \left( \frac{\lambda}{2\nu} h(\vec{r}, t) \right) - 1 \right]  \nonumber \\
&\approx&  h(\vec{r}, t) + \frac{\lambda}{4\nu} h(\vec{r}, t)^2.
\end{eqnarray}
Note that $u$ is a monotonic function of $h$, so $u$ has the same profile of maxima and minima as $h$. This new field satisfies the heat equation whose general solution is 

\begin{equation}
 u(\vec{r}, t)  \approx  \int \mathrm{d}^2r' G(\vec{r}, \vec{r'}, t) (h(\vec{r'}, 0) + \frac{\lambda}{4\nu} h(\vec{r'}, 0)^2),
 \label{eq_kpz_average}
\end{equation}
where $G(\vec{r}, \vec{r'}, t)$ denotes the Green's function.

The correlations listed in Table 2 can now be determined from the distribution of $h(\vec{r},0)$, leading to an expression for $\Delta n(t)$ that is valid over an arbitrary time span provided that $\lambda$ is small. Analytical results can be obtained for a few convenient choices of the power spectrum of $h(\vec{r}, 0)$. For example, if we take a Gaussian spectrum, $A(k)^2 \sim \exp(-k^2/2k_0^2)$, we find
\begin{equation}
\Delta n  =  \frac{\lambda}{\nu} \frac{ 16 \tau^3 (1+4\tau)^{7/2} }{ \sqrt{3\pi} (1+2\tau)^3 (1+6\tau)^4 },
\label{eq_kpz_gsn_spec}
\end{equation}
where $\tau \equiv k_0^2 \nu t$. 
The validity of this equation is illustrated in Fig.~\ref{fig_kpz}, which shows an excellent agreement between theory and numerics. The imbalance starts out at zero because the initial choice for $h$ is Gaussian. On the other hand, after long times, this expression decays back to zero. The reason for the decay is that, at long times, $u(\vec{r},t)$ involves an average over a larger and larger window, so by the central limit theorem, it starts to acquire Gaussian statistics characterized by a vanishingly small imbalance between maxima and minima. 

For early times, we can make an expansion in $t$, valid for an arbitrary power spectrum, which gives
\begin{equation}
\Delta n  =  \frac{\lambda}{\nu} \frac19 \sqrt{\frac6{\pi}} \frac1{K_2 \sqrt{K_4}} (2 K_2 K_6 - 3 K_4^2)(\nu t)^2 + O(t^3) .
\end{equation}
Note that for the Gaussian power spectrum, featured in Eq.~\eqref{eq_kpz_gsn_spec}, the second order term happens to vanish.

The agreement for the entire range of times is special for the KPZ equation: for more general equations, the agreement would break down at sufficiently late times, since the nonlinearities eventually grow exponentially; the exact transformation of the KPZ equation to a linear equation implies that the nonlinearities remain bounded.

\appendix

\section{Determination of $g(z)$}

The probability distribution $g(z)$ can be derived using a method similar to the one outlined in \cite{cite_Longuet1}. Consider a fixed point $\vec{r}$. We wish to know the probability density that at this point we have $H_G = z$ (to avoid confusion with the derivatives of $H_G$, we shall write $H$ from now on) given that it is a minimum. The conditions for this can be written in terms of derivatives of $H$, namely, $H_x = H_y = 0$, defines a critical point while $H_{xx}H_{yy} - H_{xy}^2 > 0$ and $H_{xx}+H_{yy} > 0$ distinguishes a local minimum from a saddle or maximum. 
First let us determine the joint distribution of these six variables ($H$ and its derivatives), which form a set of correlated Gaussian variables. The joint probability distribution $ p(\xi_1, \ldots, \xi_n)$ for any such set is completely determined by the correlations between the variables:

\begin{equation}
p(\xi_1, \ldots, \xi_n)  =  (2\pi \det C)^{-n/2} \exp \bigg( -\sum_{ij} C^{-1}_{ij} \xi_i \xi_j \bigg),
\label{eq_gsn_jpd}
\end{equation}
where $C_{ij} = \la \xi_i \xi_j \ra$ is the matrix of correlations between the variables. 

Correlations between $H$ and its first and second derivatives can be expressed in terms of the first three moments ($K_0,K_2,K_4$) of its amplitude spectrum.
By differentiating the Fourier expansion of $H$ we find that $\langle H_x^2\rangle=\langle H_y^2\rangle=\frac{1}{2}K_2$, and likewise that the variances of the second derivatives are proportional to $K_4$. The only variables among the six that are correlated \emph{to one another} turn out to be $H$, $H_{xx}$ and $H_{yy}$, with $\la H H_{xx} \ra = \la H H_{yy} \ra = -K_2/2$ and $\la H_{xx} H_{yy} \ra = K_4/8$. After retrieving the probability distribution, we set $H = z$, $H_x = H_y = 0$ and integrate over $H_{xx}$, $H_{yy}$ and $H_{xy}$ (over the domain defining a minimum). The Jacobian determinant $|H_{xx}H_{yy}-H_{xy}^2|$ must be added \cite{cite_Longuet1, cite_longpaper1}. The probability density we have calculated so far reflects the chance that $H_x$ and $H_y$ are close to zero at the point $\vec{r_0}$ (there is a vanishing chance that they are \emph{exactly} 0).
However, we want to find the probability of the reverse situation, that $H_x=H_y=0$ exactly at a point within a small range of $\vec{r}_0$ (since we are looking at the distribution of extrema in the plane). The ratio of the two probabilities is given by the Jacobian determinant. The final answer reads
\begin{equation}
 \begin{split}
   g(z) =\:	& \sqrt{\frac{3}{2\pi(3-2\lambda)}} \, e^{ -\frac{3z^2}{2(3-2\lambda)} } \erfc\Bigg( \sqrt{\frac{\lambda}{2(1-\lambda)(3-2\lambda)}} \, z \Bigg) \\
   		& - \sqrt{\frac{3}{2\pi}} \lambda (1-z^2) \, e^{ -\tfrac12 z^2 } \erfc\Bigg( \sqrt{\frac{\lambda}{2(1-\lambda)}} \, z \Bigg) \\
   		& - \frac1{\pi} \sqrt{3\lambda(1-\lambda)} \, z \, e^{ -\frac{z^2}{2(1-\lambda)} },
 \end{split}
 \label{eq_distr}
\end{equation}
where $\lambda = \frac{K_2^2}{K_0 K_4}$.

\section{Complex notation and cumulants}

Introducing the complex variables $z=x+iy$ and $z^*$ allows one to use isotropy to calculate the probability distribution very efficiently. The isotropy causes many of the cumulants of the distribution to vanish -- this explains why the monstar fraction is always the same for Gaussian fields when they are isotropic and why the shift of the monstar fraction depends on the nonlinearity $f(H)$ in a simple way.

The isotropy implies that a moment like $\la h_{zz} h_{zz^*z^*} \ra$ does not change when the field is rotated by an angle $\alpha$. On the other hand this rotation transforms $z \rightarrow z e^{i\alpha}$ and $z^* \rightarrow z^* e^{-i\alpha}$, which in the given example would introduce an extra factor $e^{i\alpha}$ in the moment. Since the moment cannot change, it must be vanishing. In general, a moment can only be nonzero if the number of $z$ and $z^*$ derivatives match.

In these complex variables, the definition of an umbilic is given by $h_{zz} = 0$. Monstars are distinguished from lemons and stars with the conditions
\begin{subequations}
\begin{align}
  |h_{zzz^*}|^2 - |h_{zzz}|^2						& > 0, \\
  \begin{split}
    27|h_{zzz}|^4 - |h_{zzz^*}|^4 - 18|h_{zzz}|^2 |h_{zzz^*}|^2 \qquad	& \\
    -\: 4(h_{zzz}h_{zz^*z^*}^3 + h_{z^*z^*z^*}h_{zzz^*}^3)		& > 0,
  \end{split}
\end{align}
\label{eq_monstar}
\end{subequations}
which translates in complex notation the condition derived in Ref. \cite{cite_Berry} for an umbilic to be a monstar.

First, we review why the monstar fraction is a constant for a Gaussian field $H$. Since by the conditions outlined above, the monstar fraction depends only on the joint distribution of $H_{zz}$, $H_{zzz}$, $H_{zzz^*}$ and their conjugates, and the properties of a Gaussian field are determined by covariances, there are only a few variables the monstar fraction can depend on. These are $\sigma=\la|H_{zz}|^2\ra$, $\tau=\la |H_{zzz}|^2\ra$, and $\tau'=\la |H_{zzz^*}|^2\ra$. The latter two are equal thanks to translational symmetry, and since there is no dimensionless function of $\sigma$ and $\tau$, $\alpha_M$ must be a constant.

Similar arguments can be used to show that the shift in $\alpha_M$ for the non-Gaussian $h=H+\varepsilon f(H)$ depends only on $\langle f'''(H)\rangle$ to first order. First, we need to determine all the cumulants of $h_{zz},h_{zzz}$ and $h_{zzz^*}$ up to first order in $\varepsilon$. At this order, we will show that most of the cumulants vanish. We are faced with cumulants of the form $C_n(D_1 h, D_2 h, \ldots, D_n h)$, where the operators $D_i$ represent two or three derivatives. Each operator acts on $h(\vec{r})$ at the same point $\vec{r}$. For the moment, we will consider each operator $D_i$ to act on a different point $\vec{r_i}$. This allows us to bring all the operators outside the cumulant. This yields $D_1 \ldots D_n C_n(h_1, \ldots, h_n)$, where $h_i = h(\vec{r_i})$.

Since we are only working up to first order, we can expand the cumulant as
\begin{equation}
\begin{split}
  C_n(h_1, \ldots, h_n)  =\:	& C_n(H_1, \ldots, H_n) \\
				& + \varepsilon C_n(f(H_1), H_2, \ldots, H_n) + \ldots \\
				& + \varepsilon C_n(H_1, H_2, \ldots, f(H_n)),
\end{split}
\end{equation}
which consists of one leading order term and $n$ first order terms for which we can apply
\begin{equation}
\begin{split}
  & C_n(f(H_1), H_2, \ldots, H_n) \\
  & \qquad  =  \la f^{(n-1)}(H_1) \ra \la H_1 H_2 \ra \la H_1 H_3 \ra \ldots \la H_1 H_n \ra.
\end{split}
\end{equation}
In this expression, the operators $D_2$ through $D_n$ can easily be reinserted. For the operator $D_1$ the product rule needs to be applied, which in principle gives rise to a lot of terms. Remember though that, after setting all $\vec{r_i}$ equal again, each moment can only be nonzero if the number of $z$ and $z^*$ derivatives match. This criterion kills most of the terms. For example, if we consider the cumulant $C_3(h_{zzz^*}, h_{zzz^*}, h_{z^*z^*})$, we encounter the term $\partial_{1,zzz^*} \la f''(H_1) \ra \la H_1 H_{zzz^*} \ra \la H_1 H_{z^*z^*} \ra$. In the product rule, after setting $\vec{r_1}$ equal to the other $\vec{r_i}$, the only nonzero term is $\la f''(H) \ra \la H_{z^*} H_{zzz^*} \ra \la H_{zz} H_{z^*z^*} \ra$.

We can now also see that there are no cumulants $C_n$ with $n > 4$ which are nonzero up to first order in $\varepsilon$. This is because when we apply the above recipe, we have $n-1$ moments of the form $\la H_1 D_i H_i \ra$. The operator $D_1$ has only three derivatives at most, therefore, in every term in the product rule there must be at least one moment of the form $\la H D_i H_i \ra$. However, the variables that we consider, $h_{zz}$, $h_{zzz}$, $h_{zzz^*}$ and their conjugates, all do not have the same number of $z$ and $z^*$ derivatives. That means that this moment must be zero, and hence the entire term too.


\begin{acknowledgments}
This work was supported by the Dutch Foundation for Fundamental Research on Matter (FOM).
\end{acknowledgments}

\end{document}